\documentstyle[11pt]{article}
\title{Random quantum correlations and \\ density operator distributions}
\author{Michael J. W. Hall\\Department of Theoretical Physics\\
Australian National University\\Canberra, ACT 0200, Australia\\*\\
Address from 20 December 1997 - 30 September 1998:\\
Universit\"{a}t Ulm\\
Abteilung f\"{u}r Quantenphysik\\Albert-Einstein Allee 11\\
89081 Ulm, Germany.\\
email:hall@physik.uni-ulm.de; fax:(0731) 502 3086}

\date{}
\begin{document}
\maketitle
%\newpage

\begin{abstract}
Randomly correlated ensembles of two
quantum systems are investigated, including average entanglement
entropies and probability distributions of 
Schmidt-decomposition coefficients. Maximal correlation is guaranteed
in the limit as one system
becomes infinite-dimensional.
The reduced density operator distributions
are compared with distributions induced
via the Bures and Hilbert-Schmidt metrics.\\



\end{abstract}

%\newpage
\section{Introduction} \
This Letter is primarily motivated by the following question:  what 
statistical ensemble corresponds to
{\it minimal} prior knowledge about a quantum system?  
Such an ensemble may be identified as the 
most {\it random} ensemble of possible 
states of the system.  It would provide, for example,
a natural benchmark for assessing 
how ``random'' a given evolution process is \cite{woott};
a worst-case scenario for general
schemes for extracting information about the system
\cite{jones}; and a natural unbiased measure 
over the set of possible states of the system
(which would allow one to
calculate, e.g., the average effectiveness of a general
scheme for distinguishing between quantum states \cite{braun}).
 
For the case where the system is known to 
in fact be in a {\it pure} state,
there is an obvious answer to the question. In particular, 
identifying minimal knowledge with maximal symmetry,
it is natural to require that the ensemble be invariant under 
the full group of unitary transformations (thus there is no
preferred measurement basis for extracting information).
This requirement yields a {\it unique} probability distribution over
the set of pure states of the system \cite{jones,skyora},
which has found applications in
quantum inference \cite{jones},quantum chaos \cite{woott}, 
and quantum information \cite{josza} 
\footnote{
If the average density
operator 
of the ensemble is also known, then the above-mentioned 
pure-state ensemble may be modified 
to give a corresponding maximally-random or ``Scrooge'' ensemble,
with the property of being that 
ensemble of pure states on which 
measurement yields the least possible 
information consistent with the prior knowledge \cite{josza}.}.

However, as pointed out by Wootters \cite{woott}, 
there does not appear to be 
a natural generalisation of the above ensemble when the 
restriction of pure states is removed.
Indeed, if general states described by  density operators are allowed,
the requirement of unitary invariance only implies that the
probability measure over the set of possible states is
a function of the density
operator eigenvalue spectrum alone.  
Hence a unique probability measure can be 
specified only via some further principle or restriction, 
to be motivated on physical or conceptual grounds.

In this Letter two possible approaches to the question 
are examined.  The first is
motivated by recent work of Braunstein \cite{braun}, and is 
considered in
sections 2 and 3 below.  It corresponds to assuming
that the quantum system is randomly 
correlated with a second system, where
the composite system is in a pure state.  The reduced ensemble of the
system is characterised by the distribution of Schmidt-decomposition
coefficients of the composite system, and is explicitly calculated
for the 2-dimensional case.  This further allows calculation of the 
average ``entanglement entropy'' \cite{benne} of the systems.
In the limit as the dimension of the auxilary
system becomes infinite, the systems become maximally correlated
with probability unity.

The second approach, studied in section 4 below, is more 
formal in nature.  It relies on choosing 
a metric on the space of density operators of the system; the random
ensemble then corresponds to the (normalised) volume element on this
metric space.  There are strong information-theoretic grounds for
motivating the choice of the Bures ``distinguishability'' 
metric \cite{bures}.
Moreover, for a 2-dimensional system, this metric corresponds
to the conceptually satisfying case of a maximally symmetric space,
with no preferred locations or directions in the space of
density operators.  The ensemble induced by the Hilbert-Schmidt
metric is also considered.

Finally, in Section 5 comparisons are made between 
the above two approaches
for the two-dimensional case.
It is argued that it is the second approach,
based on the Bures metric, which yields the desired 
``minimal knowledge'' ensemble in this case.

\section{Randomly correlated ensembles} \
Let quantum system $S$, with Hilbert space 
$H_{S}$, be correlated with  an
auxilary system $A$, with Hilbert space
$H_{A}$ (for example, a similar system, a measuring apparatus, or
the environment).  A general pure state 
of the composite system then has
the form
\begin{equation} \label{psi}
\mid \psi \rangle = \sum_{i=1}^{M} \sum_{j=1}^{N} 
c_{ij} \mid u_{i} \rangle \otimes \mid v_{j} \rangle   ,
\end{equation}
where $\{ \mid u_{i} \rangle \}$ and $\{ \mid v_{j} \rangle \}$
denote orthonormal bases for $H_{S}$ and $H_{A}$ respectively, and $M$
and $N$ are the respective dimensions of $H_{S}$ and $H_{A}$.

It is always possible to choose orthonormal bases 
$\{ \mid u_{i}^{*} \rangle \}$, $\{ \mid v_{j}^{*} \rangle \}$ for
$H_{S}$ and $H_{A}$ in which $\mid \psi \rangle$ has the
Schmidt-decomposition form \cite{schmi}
\begin{equation} \label{schmidt}
\mid \psi \rangle = \sum_{k=1}^{K} 
\sqrt{\lambda_{k}} \mid u_{k}^{*} \rangle \otimes
\mid v_{k}^{*} \rangle  ,
\end{equation}
where $K$$\leq$$\min(M,N)$, and the Schmidt coefficients 
$\{\lambda_{k}\}$ are non-zero and unique up
to permutations.  These coefficients are
just the (non-zero) eigenvalues of the reduced
density operators $\rho_{S} = tr_{A}[\mid \psi \rangle 
\langle \psi \mid ]$ and $\rho_{A} = tr_{S}[\psi \rangle
\langle \psi \mid ]$ of $S$ and $A$ respectively, as may be verified 
directly from Eq.(\ref{schmidt}).  The quantity
\begin{equation} \label{entang}
E_{\psi} = - \sum_{k} \lambda_{k} \log_{2} \lambda_{k}
\end{equation}
is called the ``entanglement entropy'' of the two systems 
\cite{benne}, and is
a useful measure of the degree of correlation between $S$ and $A$ 
\cite{braun,benne,barne}.

It will now be assumed that $S$ and $A$ are {\it randomly} correlated,
by which it is meant that the composite system is a member of
the maximally random pure-state ensemble discussed in the 
Introduction, described by
a uniform distribution over the pure states of $H_{S}\otimes H_{A}$.
Such correlations may arise if the composite system is ``chaotic'' in
the sense of Schack and Caves \cite{chaos}
(i.e., if its state is randomised over the Hilbert space by 
stochastic fluctuations). They are also relevant if two observers 
observe the respective
$S$ and $A$ components of an ensemble of pure composite systems (possibly
for cryptographic key generation \cite{benne}), in the case 
of minimal knowledge
about the ensemble.  Further, for $M=2$, Braunstein has used 
such randomly correlated ensembles to numerically
generate reduced density operators of $S$, to
test the average effectiveness 
of a general measurement scheme for
distinguishing between two states of $S$ \cite{braun}.

Now, for a quantum system of dimension $D$, the maximally random
pure-state ensemble over states $\{\mid\sigma\rangle\}$
of the system is described by the probability measure 
\cite{jones,skyora}
\begin{equation} \label{measure}
d\Omega_{\sigma} = K_{D} \delta (\langle \sigma 
\mid \sigma \rangle
- 1) \prod_{d=1}^D dRe\{\sigma_{d}\} dIm\{\sigma_{d}\}  ,
\end{equation}
where the $\{\sigma_{d}\}$ are the coefficients 
of $\mid\sigma\rangle$ with respect to some (arbitrary)
orthonormal basis, and the normalisation factor $K_{D}$
is given by
\begin{equation} \label{kmn}
K_{D}=(D-1)!/\pi^{D} .
\end{equation}
Hence the randomly correlated ensemble is described by 
the corresponding probability measure $d\Omega_{\psi}$ over the pure
states $\mid \psi \rangle$ of $H_{S}\otimes H_{A}$:
\begin{equation} \label{omega}
d\Omega_{\psi} = 2^{-MN}K_{MN} \delta (\langle \psi \mid \psi \rangle
- 1) \prod_{i,j} dp_{ij} d\phi_{ij} ,
\end{equation}
where the coefficients
$c_{ij}$ in Eq. (\ref{psi}) have the polar form 
$(p_{ij})^{1/2}$$\exp (i\phi_{ij})$.  

It proves useful to rewrite Eq. (\ref{omega}) via the definitions
\begin{eqnarray} \label{xdef}
x_{i} & = & \sum_{j} p_{ij} ,\\
\label{Pdef}
P_{ij} & = & p_{ij}/x_{i} ,\\
\label{alphdef}
\mid \alpha_{i} \rangle & = & \sum_{j} \sqrt{P_{ij}} 
\exp (i\phi_{ij}) \mid v_{j} \rangle ,\\
\label{alphmeas}
d\Omega_{\alpha_{i}} & = & 2^{-N}K_{N} \delta (\langle \alpha_{i} 
\mid \alpha_{i} \rangle
- 1) \prod_{j} dP_{ij} d\phi_{ij} ,
\end{eqnarray}
where $d\Omega_{\alpha_{i}}$ is the uniform measure 
over pure states $\{\mid \alpha_{i} \rangle\}$ of $H_{A}$.
In particular, if for each $i$ the variables $p_{ij}$ 
in Eq. (\ref{omega}) are transformed
to the variables $P_{i1}$,$\dots$,$P_{i,N-1}$ and $x_{i}$,
one obtains
\begin{equation} \label{trans}
\prod_{j=1}^{N} dp_{ij} = (x_{i})^{N-1} dx_{i}
\prod_{j=1}^{N-1} dP_{ij} ,
\end{equation}
where the Jacobian factor $(x_{i})^{N-1}$ is most
easily evaluated by adding the first $N-1$ rows of the 
Jacobian determinant to the last row.
Substituting Eqs. (\ref{xdef})-(\ref{trans}) into
Eq. (\ref{omega}), and multiplying by  dummy terms of the form
$\delta (\langle\alpha_{i}\mid\alpha_{i}\rangle$$-1)$
$dP_{iN}$, yields the final symmetric expression
\begin{equation} \label{finomega}
d\Omega_{\psi} = K_{MN}/(K_{N})^{M} 
\delta (\sum_{i} x_{i} - 1)
\prod_{i=1}^{M} (x_{i})^{N-1} dx_{i} d\Omega_{\alpha_{i}} 
\end{equation}
for the randomly correlated ensemble.

\section{Statistical properties} \
Since the measure $d\Omega_{\psi}$ is invariant under
unitary transformations \cite{jones,skyora}, the reduced ensemble of 
density operators $\rho_{S}$$=$$tr_{A}[\mid \psi \rangle
\langle \psi \mid ]$, corresponding to system $S$, is similarly
invariant under such transformations.  It follows that 
the distribution of
density operators $\rho_{S}$ is basis-independent,
and hence that the reduced ensemble is characterised by a 
probability distribution over
the eigenvalue spectrum of $\rho_{S}$.  As noted following
Eq. (\ref{schmidt}), this spectrum is
determined by the Schmidt-decomposition 
coefficients of $\mid \psi \rangle$,
and hence the corresponding 
probability distribution will be denoted by 
$p_{M,N}(\lambda_{1},\lambda_{2},\dots)$.  Note that the symmetry
between systems $S$ and $A$ implies that 
\begin{equation} \label{symm}
p_{M,N} \equiv p_{N,M} .
\end{equation}

To calculate the distribution 
$p_{M,N}$,
and hence such quantities such as the average entanglement entropy,
note from Eqs. (\ref{psi}) and (\ref{xdef})-(\ref{alphdef}) that
$\rho_{S}$ has the general form of an $M\times M$ matrix,
with coefficients
\begin{equation} \label{mat}
\langle u_{i} \mid \rho_{S} \mid u_{j} \rangle =
\sqrt{x_{i}x_{j}} \langle \alpha_{i} \mid \alpha_{j} \rangle
\end{equation}
with respect to the $\{\mid u_{i} \rangle\}$ basis.
Hence, if a general expression for the eigenvalue spectrum of
this matrix can be given,
$p_{M,N}(\lambda_{1},\lambda_{2},\dots )$ can be 
calculated from $d\Omega_{\psi}$ in Eq. (\ref{finomega}).  
This approach is
successfully followed below for $M=2$, while a less direct approach
allows calculation in the limit as $N \rightarrow \infty$.

Suppose then that $M=2$. The eigenvalues of $\rho_{S}$ follow from
Eq. (\ref{mat}) as
\begin{equation} \label{eig}
\lambda_{1} = \frac{1}{2} (1\pm r), 
\lambda_{2} = \frac{1}{2} (1\mp r) ,
\end{equation}
where
\begin{equation} \label{rdef}
r = [1 - 4x_{1}x_{2}(1 - \mid
\langle  \alpha_{1} \mid \alpha_{2} \rangle \mid^{2} )]^{1/2} .
\end{equation}
As shown in the Appendix, one finds 
\begin{equation}  \label{p2n}
p_{2,N}(\lambda_{1},\lambda_{2}) =
\frac{(2N-1)! \delta (\lambda_{1}+\lambda_{2}-1)}
{2 (N-2)! (N-1)!}(\lambda_{1}-\lambda_{2})^{2} 
(\lambda_{1} \lambda_{2})^{N-2} ,
\end{equation}
describing a two-dimensional system $S$ randomly correlated
with an $N$-dimensional system
$A$.  

The average entanglement entropy of 
$S$ and $A$ can be calculated from
Eqs. (\ref{entang}) and (\ref{p2n}) using standard integrals,
with the final result
\begin{equation} \label{entang2n}
\langle E_{\psi} \rangle =
\frac{\log_{2}e}{4^{N-1}} \frac{(2N-1)!}{(N-2)! (N-1)!}
\sum_{s=0}^{N-2} \left( \begin{array}{c} N-2 \\ s \end{array}
\right) \frac{(-1)^{s}}{(s+2)(2s+3)} \sum_{t=0}^{s+1}
\frac{1}{2t+1} .
\end{equation}
For the case of two randomly correlated qubits
($N=2$), this yields a value of
$(\log_{2}e)/3$$\approx$$0.481$ bits,
which is about half of the maximum possible value of $1$ bit.
In the limit as $N\rightarrow\infty$ the average entanglement
entropy monotonically approaches this maximum
(e.g., for $N=100$ the average entanglement entropy
is $0.99$ bits). Thus
maximal correlation between $S$ and $A$ is {\it guaranteed}
in this limit.

The latter result holds more generally. In fact, 
for arbitrary $M$ it can be shown
that the reduced ensemble contains only {\it one} density
operator, $M^{-1} \hat{1}$, in the limit 
$N\rightarrow\infty$. The corresponding eigenvalues (and hence the 
Schmidt-decomposition coefficients) each equal $M^{-1}$,
and hence from Eq. (\ref{entang}) the average
entanglement entropy attains its maximum value of $\log_{2}M$, 
i.e., the systems are maximally-correlated.

To show $\rho_{S}\rightarrow M^{-1} \hat{1}$, 
note that integrating over the vectors 
$\{\mid\alpha_{i}\rangle\}$ in
Eq. (\ref{finomega}) yields the marginal probability distribution
\begin{equation} \label{diag}
p(x_{1},\dots,x_{M}) = K_{MN} (K_N)^{-M} \delta
(x_{1}+\dots+x_{M} - 1) (x_{1}\dots x_{M})^{N-1}
\end{equation}
for the diagonal elements of
$\rho_{S}$ in the $\{\mid u_{i}\rangle\}$ basis.  Using
Stirling's approximation for $n!$ in Eq. (\ref{kmn}), it
follows that this distribution vanishes everywhere in the
limit $N\rightarrow\infty$, except for the case
$x_{i}$$\equiv$$1/M$ for all $i$.  Since the reduced 
ensemble is invariant under unitary transformations, the 
diagonal elements of $\rho_{S}$ in this limit 
are therefore equal to
$1/M$ relative to {\it any} basis.  
Choosing a basis in which $\rho_{S}$
is diagonal  gives $\rho_{S}$
$\equiv$$M^{-1} \hat{1}$ as claimed.

The above results imply that the limit
$N\rightarrow\infty$ does {\it not} yield a particularly
``random'' reduced ensemble of density operators for the
system -- indeed, it
gives an ensemble with only one member. Thus, for example,
the numerical evaluation of averages over the reduced ensemble
in Section 7 of \cite{braun} for $M=2$,
to test the average effectiveness
of a particular measurement scheme, is of most value
for the maximally random case $N=2$.

The fact that ``randomness'' is in fact decreased 
as the dimension of the auxilary system increases
suggests that a potential
candidate for the minimal-knowledge ensemble discussed in 
the Introduction is the reduced ensemble corresponding 
to $N$$=$$M$, with corresponding eigenvalue distribution $p_{M,M}$
(choosing $N$ {\it less} than $M$ would unduly restrict the
ensemble to density operators with $M-N$ zero eigenvalues).
However, other potential candidates may be generated by
a second approach, as shown in the next Section.

\section{Metric-induced ensembles} \

An ensemble of general states of a quantum system
is in general described by a probability measure over
the density operators of the system.  Given that
probability measures transform in the same way
as volume elements under co-ordinate transformations,
and that volume elements are in general properties
of metric spaces, this suggests that the distribution of
density operators corresponding to a ``minimal-knowledge''
ensemble may be obtained from the normalised 
volume element induced by
some natural metric on the space of density operators.

A metric of particular interest is the Bures metric 
\cite{bures}, 
where the infinitesimal distance element between
two states $\rho$ and $\rho +$$\delta\rho$ is given by
\cite{hubner}
\begin{equation} \label{bures}
(ds_{B})^{2} = 2 \sum_{j,k} (\lambda_{j}+\lambda_{k})^{-1}
\mid\langle j\mid\delta\rho\mid k\rangle\mid^{2} ,
\end{equation}
where $\rho$ is diagonal in the orthonormal basis 
$\{\mid j\rangle\}$ with eigenvalues $\{\lambda_{j}\}$.
This metric provides a  unitarily-invariant measure 
for distinguishing between two quantum states, and
has been strongly motivated as physically 
relevant both on measurement
\cite{brauncaves} and statistical \cite{braun,josza2}
grounds.
 
To calculate the volume element corresponding to the Bures metric,
it is useful to decompose $\rho +$$\delta\rho$ 
as an infinitesimal shift in the eigenvalues of
$\rho$ followed by an infinitesimal unitary transformation:
\begin{eqnarray}
\rho + \delta\rho & = & (\hat{1} + \delta U) (\rho +
\delta\Lambda) (\hat{1} + \delta U)^{\dagger} \nonumber\\
\label{inf}
& = & \rho + \delta\Lambda + [\delta U, \rho] ,
\end{eqnarray}
where $\langle j\mid\delta\Lambda\mid k\rangle =$
$\delta_{jk} d\lambda_{j}$, and $(\delta U)^{\dagger} =$
$-\delta U$ follows from unitarity.
Note moreover that the infinitesimal generator $\delta U$
can generally be decomposed as
\begin{equation} \label{gen}
\delta U = \sum_{j\leq k} [(dx_{jk} + i dy_{jk})
\mid j\rangle\langle k\mid - h.c. ]
\end{equation}
where $dx_{jk}$ and $dy_{jk}$ are real, and $h.c.$ denotes
the Hermitian conjugate of the expression preceding it.

Substitution of Eqs. (\ref{inf}) and (\ref{gen}) into
Eq. (\ref{bures}) yields
\begin{equation}
(ds_B)^{2} =  \sum_{j} \frac{(d\lambda_{j})^2}{\lambda_{j}}
+ 4 \sum_{j<k} \frac{(\lambda_{j}-\lambda_{k})^2}
{\lambda_{j}+\lambda_{k}} [(dx_{jk})^2 + (dy_{jk})^2] ,
\end{equation}
from which one immediately extracts the volume element
\begin{equation} \label{vol}
dV_{B} = \frac{d\lambda_{1}\dots d\lambda_{M}}
{(\lambda_{1}\dots\lambda_{M})^{1/2}} \prod_{j<k}
4 \frac{(\lambda_{j}-\lambda_{k})^2}
{\lambda_{j}+\lambda_{k}} dx_{jk} dy_{jk} .
\end{equation}
Normalising $dV_{B}$ yields the desired probability
distribution over the space of density operators.

Since the metric is invariant under unitary
transformations, the corresponding ensemble
is characterised by  the marginal probability distribution
$p_{B}(\lambda_{1},$$\dots,$$\lambda_{M})$ over the eigenvalue
spectrum of the density operators describing the system
(see also Section 3).  This distribution can be obtained from Eq.
(\ref{vol}) by integrating over the (compact) space of unitary
transformations (parametrised by $\{ x_{jk}, y_{jk}\}$),
and normalising, to give
\begin{equation} \label{pb}
p_B(\lambda_{1},\dots ,\lambda_{M}) =
C_{M} \frac{\delta (\lambda_{1}+\dots +\lambda_{M} - 1)}
{(\lambda_{1}\dots \lambda_{M})^{1/2}} \prod_{j<k}
\frac{(\lambda_{j}-\lambda_{k})^2}{\lambda_{j}+\lambda_{k}} ,
\end{equation}
where $C_{M}$ is a normalisation constant, and the condition
$tr[\rho ]=1$ has been made explicit.

Eq. (\ref{pb}) will be compared with 
Eq. (\ref{p2n}) in the following
Section for the case $M=2$.  This Section is concluded by
noting that in principle there are many 
possible choices of metric, each leading
to a possible ``random'' ensemble.  One particularly simple
choice is the Hilbert-Schmidt metric, with infinitesimal distance
element
\begin{equation} \label{hilbert}
(ds_{HS})^2 = tr[(\delta\rho )^{2}] .
\end{equation}
Following essentially the same procedure as above for the Bures
metric (where the trace is evaluated in the $\{\mid j\rangle\}$
basis), one finds the corresponding probability distribution
\begin{equation} \label{phs}
p_{HS}(\lambda_{1},\dots ,\lambda_{M}) =
{C'}_{M} \delta (\lambda_{1}+\dots +\lambda_{M} - 1)
\prod_{j<k} (\lambda_{j}-\lambda_{k})^2
\end{equation}
for the density operator eigenvalue spectrum.  This is also
considered in the following Section for the case $M=$$2$.

\section{Two-dimensional comparisons} \
The states $\rho$ of a two-dimensional system 
may be parametrised in the Bloch representation as 
\begin{equation} \label{bloch}
\rho = \frac{1}{2} (1 + {\bf \sigma\cdot r}) ,
\end{equation}
where ${\bf \sigma}$ is the 3-vector of Pauli matrices and 
${\bf r}$ is a 3-vector of modulus $r\leq 1$.  The eigenvalues
of $\rho$ are related to $r$ as per Eq. (\ref{eig}).  

A distribution over $\rho$ may therefore be written as a
distribution over ${\bf r}$.  
Moreover, since unitary transformations of $\rho$
correspond to rotations of ${\bf r}$, it follows that 
distributions corresponding to unitarily-invariant 
ensembles depend only
on the modulus $r$, being uniform with respect to direction.
Hence the distributions over ${\bf r}$ corresponding to
eigenvalue distributions Eqs. (\ref{p2n}) (with $N=2$), 
(\ref{pb}) and (\ref{phs}) are given respectively by
\begin{eqnarray} \label{p22}
p_{2,2}({\bf r}) & = & 3/(4\pi) ,\\
\label{pb2}
p_{B}({\bf r}) & = & (4/\pi) (1-r^{2})^{-1/2} ,\\
\label{phs2}
p_{HS}({\bf r}) & = & 3/(4\pi) .
\end{eqnarray}

It is seen that the first  and third distributions are uniform over
the unit 3-ball, while the distribution corresponding to the Bures
metric is sharply peaked at the surface of the ball (corresponding 
to pure states of the system).  This raises the question of which 
is the more ``random''?  I shall argue here for the latter, due to
its greater symmetry.

In particular, the Bures metric for a two-dimensional system 
corresponds to the surface of a unit 4-ball \cite{hubner},
i.e., to the maximally symmetric
3-dimensional space of positive curvature \cite{weinberg}
(and may be recognised as the spatial part of the Robertson-Walker
metric in general relativity \cite{weinberg}).  This space is
homogenous and isotropic, and hence the 
Bures metric does not distinguish
a preferred location or direction in the space
of density operators.  Indeed, as well as rotational symmetry
in Bloch co-ordinates (corresponding to unitary invariance), 
the metric has a further set
of symmetries generated by the infinitesimal transformations
\cite{weinberg}
\begin{equation} \label{quasi}
{\bf r}\rightarrow{\bf r}+\epsilon (1-r^{2})^{1/2}{\bf a}
\end{equation}
(where ${\bf a}$ is an arbitrary 3-vector).

Taking the viewpoint that maximal randomness corresponds to an 
ensemble with maximal symmetry, it follows that the
distribution of Eq. (\ref{pb2}), in corresponding to a maximally
symmetric space, is in fact more ``random'' than the distributions of 
Eqs. (\ref{p22}) and (\ref{phs2}).
This strongly suggests, at least for
two-dimensional quantum systems, that the minimal-knowledge
ensemble discussed in the Introduction is the one induced by
the Bures metric.

Finally, note that the existence of various candidates for
the minimal-knowledge ensemble
discussed in the Introduction
begs the question as to whether there
exists some natural physical process for generating ensembles of
quantum systems,
which can be identified with maximal randomness.  This would allow
experimental determination of the minimal-knowledge ensemble.
This is, however, beyond the scope of this Letter.

{\bf Acknowledgement}\\
I thank Sam Braunstein for useful discussions, and
Robyn Hall for partial financial support.

\appendix
{\bf Appendix}\

To derive Eq. (\ref{p2n}), note from Eqs. (\ref{eig}) and 
(\ref{rdef}) that the eigenvalue distribution
can be calculated if the joint distribution of the
variables 
\begin{equation} \label{xydef}
X=x_{1}, Y=\mid\langle\alpha_{1}\mid\alpha_{2}\rangle\mid^{2}
\end{equation}
is known. From Eq. (\ref{finomega})
the statistics of $X$ and $Y$ are independent, and hence 
this joint distribution has the factored form
\begin{equation} \label{pxy}
p(X,Y) = p(X) q(Y) .
\end{equation}

From Eqs. (\ref{finomega})  and (\ref{xydef}) one 
immediately has 
\begin{equation} \label{px}
p(X) = K_{2N} (K_{N})^{-2} [X(1-X)]^{N-1} .
\end{equation}
To find $q(Y)$, fix an orthonormal basis
$\{\mid v_{j}\rangle\}$ in $H_{A}$, and for a given unit
vector $\mid\alpha_{1}\rangle$ let $U$ be a unitary
transformation which maps $\mid\alpha_{1}\rangle$
to $\mid v_{1}\rangle$ and define
$\mid\beta\rangle =$$U \mid\alpha_{2}\rangle$.  Thus
\begin{equation}
Y=\mid\langle\alpha_{1}\mid U^{\dagger}U
\mid\alpha_{2}\rangle\mid^{2}=
\mid\langle v_{1}\mid\beta\rangle\mid^{2} .
\end{equation}

Since $d\Omega_{\alpha_{2}}$ is invariant under unitary
transformations, writing 
$\langle v_{j}\mid\beta\rangle$$=$$(w_{j})^{1/2}$
$\exp(i\theta_{j})$ yields
\begin{eqnarray} 
d\Omega_{\alpha_{2}}& = &d\Omega_{\beta} \nonumber\\
& = &
2^{-N}K_{N}\delta (\langle\beta\mid\beta\rangle -1)
\prod_{j} dw_{j} d\theta_{j} \nonumber\\
& = & 2^{-N}K_{N}\delta (\sum_{j=2}^{N}
w_{j} -(1-Y)) dY d\theta_{1}
\prod_{j=2}^{N} dw_{j} d\theta_{j} \nonumber\\
& = & 2^{-N}K_{N}\delta (\sum_{j=2}^{N}W_{j} - 1)
(1-Y)^{N-2} dY d\theta_{1} \prod_{j=2}^{N}
dW_{j} d\theta_{j} \nonumber\\
& = & 2^{-1} (K_{N}/K_{N-1}) (1-Y)^{N-2} dY
d\theta_{1} d\Omega_{\gamma} ,
\end{eqnarray}
where $W_{j}=$$w_{j}$$/(1-Y)$, and $d\Omega_{\gamma}$
is the uniform measure over pure states of the
$(N-1)$-dimensional space spanned by $\{\mid v_{2}
\rangle$$\dots$$\mid v_{N}\rangle\}$.  Multiplying
this expression by $d\Omega_{\alpha_{1}}$ and integrating
over all variables except $Y$ then gives 
\begin{equation} \label{qy}
q(Y) = \pi(K_{N}/K_{N-1}) (1-Y)^{N-2} .
\end{equation}

Finally, substituting Eqs. (\ref{kmn}), 
(\ref{px}) and (\ref{qy}) into
Eq. (\ref{pxy}), the
marginal distribution of $r$ in Eq. (\ref{rdef}) 
can be calculated as
\begin{equation} \label{pr}
p(r) = \frac{1}{2}\frac{(2N-1)!}{(N-1)!(N-2)!}r^{2}
\left(\frac{1-r^{2}}{4}\right)^{N-2} ,
\end{equation}
which with Eq. (\ref{eig}) immediately yields Eq. (\ref{p2n}).

%\newpage

\newpage
{\bf ADDITIONAL NOTES}
\\

Equation (17) is a special case of Theorem 3, Sec.III of S. Lloyd
and H. Pagels, in Ann. Phys. (NY) 188 (1988) 186.   Taking $M\leq N$
without loss of generality (see Equation (13)), one has the general
formula
\begin{equation}
p_{M,N}(\lambda_1, \dots ,\lambda_M) = C_{M,N} \delta (\sum \lambda_m
- 1) [\prod_{m<n} (\lambda_m - \lambda_n)^2 ] \, [\prod_k 
(\lambda_k )^{N-M} ]  ,
\end{equation}
where $C_{M,N}$ is a normalisation constant.
\\

Equation (18) can be simplified and generalised to calculate the
average entanglement entropy for all values of $M$ and $N$, using
a formula conjectured by Don Page and elegantly proved by
S. Sen in Phys. Rev. Lett. 77 (1996) 1-3.  Again taking $M\leq N$,
one has
\begin{equation}
\langle E_{\psi}\rangle_{M,N} = \sum_{m=N+1}^{MN}
\frac{1}{m} - \frac{M-1}{2N} .
\end{equation}
\\

As noted in the published version of the present paper (Phys. Lett. A
242 (1998) 123-129), a recent related preprint \cite{slater} seeks to
determine the ``maximally noninformative'' ensemble for a 
two-dimensional quantum system.  This is essentially a different concept
from the "maximally random" ensemble sought here (and identified as 
corresponding to the Bures volume measure in the 2-D case).  Indeed, 
intuitively one would expect an ensemble consisting solely of the
density operator proportional to the density operator to be least
"informative", as (i) no Shannon information can be gained by 
measurement on such an ensemble; and (ii) no preferred basis can be
singled out by measurement.
\end{document}